# Violation of the principle of Complementarity, and its implications

Shahriar S. Afshar*[a,b]
[a]Dept. of Physics, Harvard Univ., Cambridge, MA, USA 02138
[b]Dept. of Physics and Astronomy, Rowan Univ., 201 Mullica Hill Rd., Glassboro, NJ, USA 08028

**ABSTRACT**

Bohr's principle of complementarity predicts that in a *welcher weg* ("which-way") experiment, obtaining fully visible interference pattern should lead to the destruction of the path knowledge. Here I report a failure for this prediction in an optical interferometry experiment. Coherent laser light is passed through a dual pinhole and allowed to go through a converging lens, which forms well-resolved images of the respective pinholes, providing complete path knowledge. A series of thin wires are then placed at previously measured positions corresponding to the dark fringes of the interference pattern upstream of the lens. No reduction in the resolution and total radiant flux of either image is found in direct disagreement with the predictions of the principle of complementarity. In this paper, a critique of the current measurement theory is offered, and a novel nonperturbative technique for ensemble properties is introduced. Also, another version of this experiment without an imaging lens is suggested, and some of the implications of the violation of complementarity for another suggested experiment to investigate the nature of the photon and its "empty wave" is briefly discussed.

**Keywords:** Photon, complementarity, wave-particle duality, *welcher weg*, which-way experiments, parametric down-conversion, Afshar experiment, measurement theory, empty wave, wavefunction collapse

## 1. INTRODUCTION

The wave-particle duality has been at the heart of quantum mechanics since its inception. The celebrated Bohr-Einstein debate revolved around this issue and was the starting point for many illuminating experiments conducted during the past few decades. Einstein believed that one could confirm both wave-like and particle-like behaviors in the same interferometry experiment. Using a movable double-slit arrangement, he argued that it should be possible to obtain *welcher-Weg* or which-way information (WWI) for an electron landing on a bright fringe of an interference pattern (IP) "to decide through which of the two slits the electron had passed".[1] Although Einstein ultimately failed to achieve this goal, his logical consistency argument (LCA) was the initial motivation behind Bohr's Principle of Complementarity (PC).[1] The general formulation of LCA, in the context of the double-slit experiment, could read as follows:
(I) Perfectly visible IP implies that the quantum passed through *both* slits (sharp wave-like behavior).
(II) Complete WWI implies that the quantum passed through only *one* of the slits (sharp particle-like behavior).
(III) Satisfaction of both (I) and (II) in a *single* experimental setup is a logical impossibility, since (I) and (II) are mutually exclusive logical inferences. Bohr famously avoided the logical impasse mentioned in (III) by applying Heisenberg's uncertainty principle to the experimental setup,[2] showing that under any *particular* experimental configuration one can only achieve (I) or (II), and *never* both. In Bohr's own words: "…we are presented with a choice of *either* tracing the path of the particle, *or* observing interference effects…we have to do with a typical example of how the complementary phenomena appear under *mutually exclusive* experimental arrangements".[1] Several recent experiments,[3-9] however, suggest independence of the interferometric complementarity from the uncertainty principle; hence, we shall only discuss the limitations of PC in this paper. A quantitative formulation for which-way detection has been developed on the basis of theoretical[10-15] and experimental [9, 16-19] investigations of PC during the past two decades, leading to a wave-particle duality relation covering both sharp and intermediate stages expressed as:

$$V^2 + K^2 \leq 1 , \tag{1}$$

where the two complementary measurements are $0 \leq V \leq 1$, the visibility or contrast of the IP, and $0 \leq K \leq 1$ the which-way knowledge corresponding to WWI. The visibility is given by

$$V = (I_{max} - I_{min})/(I_{max} + I_{min}) , \tag{2}$$

where $I_{max}$ is the maximum intensity of a bright fringe and $I_{min}$ is the minimum intensity of the adjacent dark fringe, so that $V=1$ when the fringes are perfectly visible (sharp wave-like behavior), and $V=0$ when there is no discernible IP. By analogy, for the which-way knowledge $K_1 = (I_1 - I_2)/(I_1 + I_2)$, so $K=1$ when the WWI is fully obtained (sharp particle-like behavior), and $K=0$ when the origin of the quantum cannot be distinguished.

It is noteworthy to mention that quantum mechanics does not forbid the presence of *non-complementary* wave and particle behaviors in the same experimental setup. What is forbidden is the presence of *sharp complementary* wave and particle behaviors in the same experiment. Such complementary observables are those whose projection operators *do not commute*.[20]

In this paper we shall only investigate sharp complementary wave and particle behaviors explicitly forbidden by PC in the same experiment. Therefore, intermediate conditions, where $0 < V < 1$, and $0 < K < 1$ shall not be covered. We assume full validity for quantum mechanical formalism, and make use of it to test the predictions of PC as a particular interpretation of quantum mechanics. Finally, although in our experiments we have not used a coherent *single-photon* source, it is expected that exactly the same results would be obtained if such a source is used.

## 2. CONVENTIONAL MEASUREMENTS OF COMPLEMENTARY OBSERVABLES

### 2.1 A modern version of the principle of complementarity

We can take advantage of the recent developments in the debate over the PC to update the definition of interferometric complementarity. Based on Eq. (1) a *modern* version of the orthodox PC-the contemporary principle of complementarity (CPC)-can be formulated as follows:

> In any *particular* experimental arrangement,
> (i) If $V=1$, then $K=0$.
> (ii) If $K=1$, then $V=0$.

It is clear from CPC (i) that in any *welcher weg* experiment, obtaining full visibility for the IP should lead to a complete loss of the WWI for the quanta. Let us pay homage to orthodoxy by applying its tenets to two experiments.

### 2.2 Destructive measurement of IP visibility

In the first experiment, we test the validity of CPC(i) in a *conventional* manner. As shown in Figure 1(a), coherent and highly stable laser light of wavelength $\lambda = 650$ nm impinges upon a *dual pinhole* with a center-to-center distance of $a$ =2000 μm and pinhole diameters of $b = 250$ μm. Two diffracted beams represented by wave functions $\Psi_1$ and $\Psi_2$ emerge. The overlapping diffraction patterns of the beams caused by the corresponding pinholes are apodized (see Appendix A,) by passing the light through an aperture stop (AS) permitting only the maximal Airy disks of radius $s = 10.4$ mm to pass, thus eliminating higher order diffraction rings. A photosensitive surface is placed at plane $\sigma_1$ at a distance $l = 400$ cm from the dual pinhole, and a fully visible IP ($V=1$), with peak-to-peak distance of $u = 1.4$ mm for the consecutive fringes, is observed as shown in Figure 1(b).

Assuming that $\Psi_1$ and $\Psi_2$ are the *apodized* wave functions, the probability density, or its classical equivalent, the irradiance, for the *coherent* superposition state $\psi_{12} = \psi_1 + \psi_2$, is given by

$$I_{12} = |\psi_{12}|^2 = |\psi_1|^2 + |\psi_2|^2 + \Gamma , \qquad (3)$$

where $\Gamma = \psi_1^* \psi_2 + \psi_1 \psi_2^*$ is the usual interference term. It is clear that observing the IP in this configuration leads to a complete loss of WWI, because the photosensitive surface at $\sigma_1$ *destructively* absorbs all of the incoming light and no further analysis can take place, hence $K=0$. Here, in *conformity* with Eq. 1 the complementary measurements are $V=1$, and $K=0$.

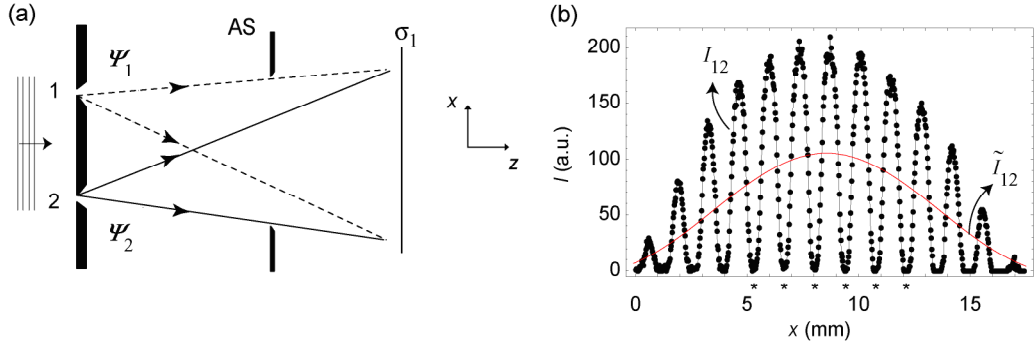

**Figure 1.** (a) Laser light impinges upon a dual pinhole and two diffracted beams $\Psi_1$ and $\Psi_2$ emerge. The beams are apodized by an aperture stop AS. (b) The interference pattern $I_{12}$ is observed at plane $\sigma_1$. Here $V=1$, and $K=0$. The red curve shows the theoretical decoherent irradiance profile $\tilde{I}_{12}$. The irradiance is measured in arbitrary units a.u. of grey-level intensity.

For comparison, the red curve shown in Figure 1(b) depicts the theoretical irradiance profile for the case $V=0$, where

$$\tilde{I}_{12} = |\psi_1|^2 + |\psi_2|^2 \qquad (4)$$

is the irradiance for the *decoherent* state, which clearly lacks any interference fringes.

### 2.3 Destructive measurement of which-way information

The application of a converging lens for which-way detection has a long history and is already implicit in the classic "Heisenberg's microscope" proof of the uncertainty principle, where the spatial resolution of the lens $\Delta x$, enters directly into the uncertainty relation $\Delta p_x \cdot \Delta x \sim h$.[2, 21, 22] Wheeler has used the lens explicitly for which-way detection in a proposed *welcher weg* experiment[23], such that photons registered at each image of the two slits are assumed to have passed through the corresponding slit, thus providing WWI.

In the second experiment, as shown in Figure 2(a), we remove the photosensitive surface at $\sigma_1$, and allow the light to pass through a suitable converging lens (L), here, with a focal length $f = 100$ cm and effective diameter of $d = 30$ mm, placed at a distance $p = 420$ cm from the pinholes, which then forms two well-resolved images (1´ and 2´) of the corresponding pinholes (1 and 2) at the image plane $\sigma_2$ at a distance of $q = 138$ cm from the lens. The image data collected at $\sigma_2$ is shown in Figure 2(b) in black. The theoretical spatial resolution of the lens in this experiment is $R \approx 30$ μm, which matches well with the observation. Less than $10^{-6}$ of the peak value irradiance from either image is found to enter the other channel, essentially providing $K=1$. For comparison, the red curve in Figure 2(b) shows the theoretical irradiance profile for a $K=0$ case (no WWI,) where a single unresolved peak instead of the two well separated peaks would be observed.

Again in this experiment, the photons are destructively detected at $\sigma_2$, and no further analysis can take place *afterwards*. However, Eq. 1 in conjunction with LCA(III) predicts a visibility of $V=0$ for the IP in this experiment, which entails a decoherent state for the two wave functions $\Psi_1$ and $\Psi_2$ at $\sigma_1$ with a corresponding *decoherent* irradiance distribution $\tilde{I}_{12} = |\psi_1|^2 + |\psi_2|^2$ as shown in Figure 1(b). In contrast to $I_{12}$, in this case the resulting irradiance $\tilde{I}_{12}$ lacks the interference term $\Gamma$.

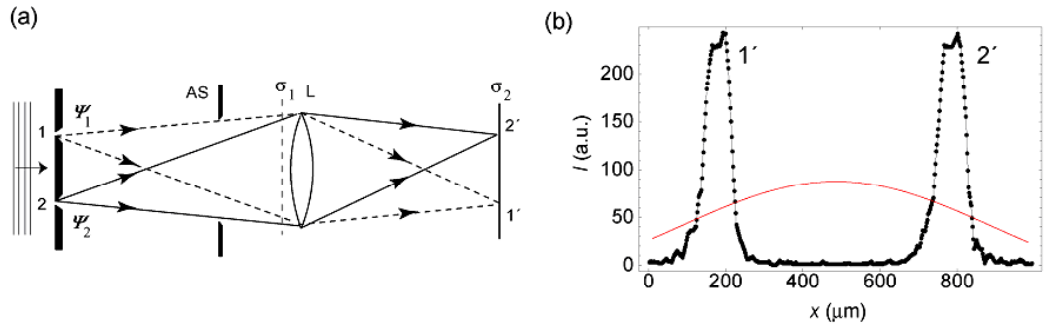

**Figure 2.** (a) A converging Lens L placed in front of $\sigma_1$ produced two well-resolved images of the pinholes. (b) The irradiance profile of the images 1´ and 2´. The photons landing in 1´ originate in pinhole 1, and those landing in 2´ originate in pinhole 2. Here, $V=0$, and $K=1$. The red curve shows a theoretical irradiance profile for a $K=0$ case.

In this experiment, the decoherence of the wave functions *prior* to entering the lens is a counter-intuitive conclusion dictated by PC, as it implies that the potential future act of obtaining WWI (the detection of the pinhole images at $\sigma_2$) leads to the loss of the IP at an *earlier* stage (at $\sigma_1$) in a non-local manner. As Feynman puts it, this situation "has in it the heart of quantum mechanics" and "contains the only mystery" of the theory.[24]

## 3. THEORETICAL DIGRESSION: MEASUREMENT THEORY REVISITED

### 3.1 Critique of the orthodox concept of "measurement"
Before we discuss the main experiment, let us momentarily take an uncustomary digression to theory to elucidate the motivation behind the experiment. Measurement in general, can be defined as *a physical process by which quantitative knowledge is obtained about a particular property of the entity under the study*. Most orthodox measurements of quantum systems involve the interaction of a microscopic quantum particle with a macroscopic classical measuring apparatus, which inevitably leads to an *irreversible* and *destructive* change in the property we want to measure. For instance, the energy of a particle can be measured by bringing it to a halt in a scintillator. This process irreversibly "destroys" the particle's energy, i.e. the particle no longer carries the initial energy after the measurement process. Although in the so-called quantum nondemolition measurements we can preserve a particular property after successive measurements, this is achieved at the expense of introducing *irreversible* perturbation to the particle's other physical properties. What these types of destructive measurements have in common is that they are performed at the level of a *single particle* and *lead to an irreversible change in the final quantum state of the detector*. It is indeed *impossible* to obtain quantitative knowledge about a particular physical property of a single particle in a non-destructive and non-perturbative manner. Unfortunately, in his reasoning for the necessity of the principle of complementarity, Bohr erroneously applies destructive measurement schemes for establishing the wave-like behavior of photons in a *welcher weg* experiment, as discussed in section 2.2.[1] However, as we shall demonstrate in the next section, the measurement of a *multi-particle* or *ensemble* property *need not be destructive*.

### 3.2 Coherence and wave-like behavior
Formation of an IP is aptly considered as evidence for coherent wave-like behavior of quantum particles. However, whereas in classical electromagnetism a *continuous* IP would be formed no matter how weak the source, in contrast quantum mechanics disallows such a state due to the fact that upon arriving at the observation plane, each quantum produces only a single dot. Figures 3(a-c) show the theoretical buildup of an IP from a coherent single-photon source over progressively extended periods of time, with 30, 300, and 3000 photons registered respectively. For comparison, Figures 3(d-f) show the decoherent photon distribution of the same number of photons respectively. It is *impossible* from the data in Figures 3(a) and 3(d), with only 30 photons registered, to discern which of the two show a coherent distribution (i.e. an IP) or a decoherent one. It is only as larger and larger numbers of photons arrive that one can recognize the lack or presence of an IP. In other words, ***evidence for coherent wave-like behavior is not a single-particle property, but an ensemble or multi-particle property***.

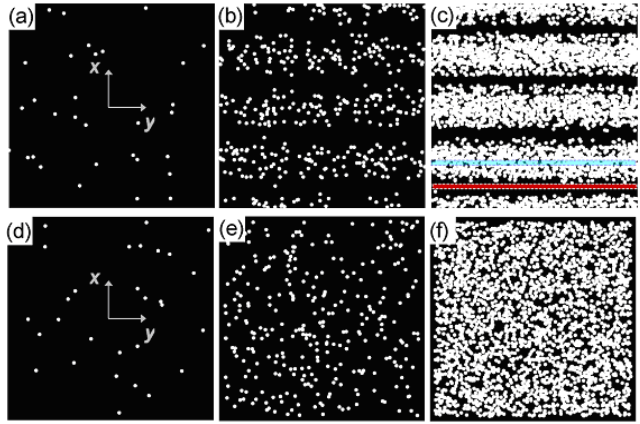

**Figure 3.** The interference pattern produced by a single-photon source with (a) 30, (b) 300, and (c) 3000 photons registered. In contrast, the decoherent distribution of (d) 30, (e) 300, and (f) 3000 photons lacks the dark fringes.

In contrast to single-particle properties such as the arrival of a single photon at a particular pinhole image, which immediately provides WWI as discussed in section 2.3, evidence for coherence *essentially* involves multiple measurements. The other important feature of coherent behavior is that there exist "forbidden" regions in space corresponding to the dark fringes, where no photons can be found. This avoidance of the dark fringe region is essential for the definition of an IP and its visibility

### 3.3 Nondestructive measurement of IP visibility
The conventional method of obtaining the visibility of an IP involves two separate measurements:

1. Destructive measurement of the maximum radiant flux at a bright fringe in order to obtain $I_{max}$.

2. Destructive measurement of the minimum radiant flux at a dark fringe in order to obtain $I_{min}$.

By substituting the values for $I_{max}$ and $I_{min}$ in Eq. (2), $V = (I_{max} - I_{min})/(I_{max} + I_{min})$, the visibility is calculated. The above process is necessary *if* $V<1$, however, if the IP is perfectly visible ($V=1$), then step 1 would be *entirely* superfluous. This is because in a perfectly visible IP, $I_{min} = 0$, and under such a condition, Eq. (2) is reduced to $V = \frac{I_{max}}{I_{max}} \equiv 1$, regardless of the actual value of $I_{max}$. *Therefore, as long as the total radiant flux of the dual pinhole output is nonzero (thus ensuring $I_{max} \neq 0$), all we need to establish perfect visibility is to determine $I_{min} = 0$.*

We can obtain $I_{min} = 0$ in two different ways: (i) by directly measuring the flux by placing a very thin detector array at the dark fringe, making sure it does not obstruct the bright fringes, or (ii) by placing an opaque obstacle such as a thin wire at the middle of a dark fringe and comparing the total radiant flux before and after the obstacle. Due to the technical impracticality of method (i), in our experiment, we opt for method (ii).

Figure 4(a) shows the schematics of method (ii) where the wire is shown as a small dark disk in the cross-section view, and $\sigma_0$ and $\sigma_1$ are parallel planes immediately before and after the wire. Assuming a coherent behavior, if we denote the distance between the centers of the pinholes as *a*, the diameter of the pinholes as *b*, the distance between the dual pinholes and $\sigma_0$ as *l*, and the wavelength of the laser as $\lambda$, then the IP is bounded within an Airy disk of radius

$$s = 3.833 \, l \, \lambda/b, \qquad (5)$$

and the distance between the peaks of each neighbouring bright fringe within the disk is

$$u = l \, \lambda/a. \qquad (6)$$

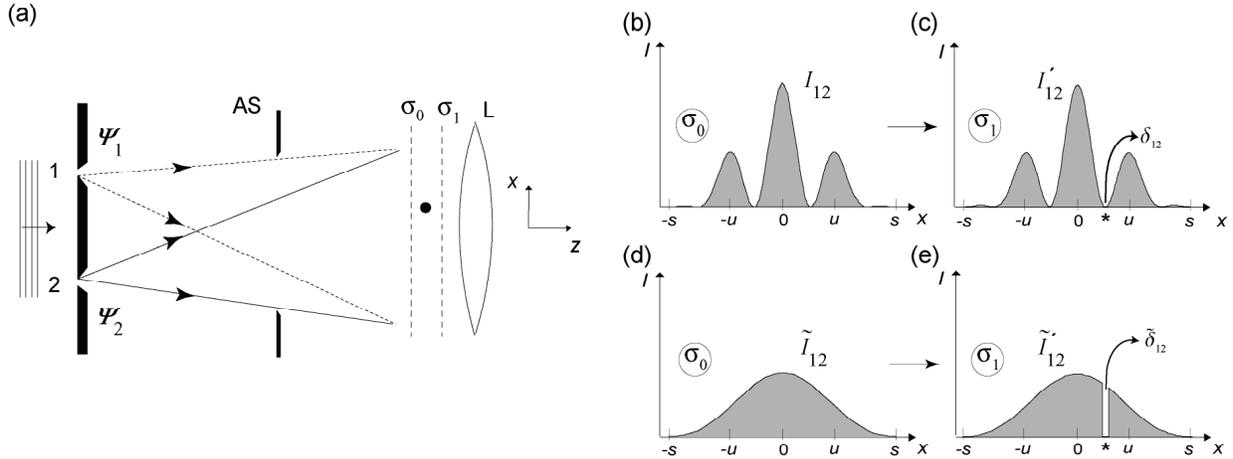

**Figure 4.** The effect of an opaque obstacle placed at the dark fringe of an interference pattern. (a) The planes $\sigma_0$ and $\sigma_1$ are located immediately before and after the obstacle, which is a wire shown as the small black disk. The irradiance profile $I_{12}$ of the coherent superposition state $|\psi_{12}\rangle$, at (b) plane $\sigma_0$, and (c) plane $\sigma_1$. The irradiance profile $\tilde{I}_{12}$ of a decoherent state, at (d) plane $\sigma_0$, and (e) plane $\sigma_1$.

The *coherent* irradiance is given by

$$I_{12} = |\psi_{12}|^2 = [2 \cos\alpha \, J_1(\beta)/\beta]^2, \tag{7}$$

$$\alpha = \pi x/u, \tag{8}$$

$$\beta = 1.22 \, \pi x/s, \tag{9}$$

and $J_1(\beta)$ is the Bessel function of first order and first kind.[25] For clarity, we have selected an IP with three bright fringes as shown in Figure 4(b). Here we assume that the thickness of the wire is $e = u/10$ and is placed at the position $x = u/2$, in the middle of the right centermost dark fringe shown as an asterisk in Figure 4(c) depicting the irradiance $I'_{12}$ at $\sigma_1$ immediately after the wire. It is clear that for the coherent case, the wire does not reduce the transmitted light appreciably, since it receives virtually no incident light such that

$$\int_{-s}^{s} I_{12} \, dx = \int_{-s}^{s} I'_{12} \, dx + \delta_{12}, \tag{10}$$

$$\delta_{12} = \int_{x_1}^{x_2} |\psi_{12}|^2 \, dx \approx 0, \tag{11}$$

where $x_1 = (u - e)/2$ and $x_2 = (u + e)/2$.

Therefore, denoting $\Phi = \|\psi\|^2 = \int_{-s}^{s} |\psi|^2 dx = \int_{-s}^{s} I \, dx$ for the total radiant flux (see A.4) we can rewrite Eq. (9) as

$$\Phi_{12} = \Phi'_{12} + \delta_{12} . \tag{12}$$

In contrast, the situation for a *decoherent* distribution, where $V=0$ is quite different. As shown in Fig. 4(d), the decoherent irradiance

$$\tilde{I}_{12} = 2\,[J_1(\beta)/\beta]^2, \tag{13}$$

also bound within the same Airy disk as the coherent state,[25] suffers a reduction in total radiant flux of

$$\tilde{\delta}_{12} = \int_{x_1}^{x_2} \tilde{I}_{12} \, dx \neq 0 . \tag{14}$$

Therefore,

$$\tilde{\Phi}_{12} = \tilde{\Phi}'_{12} + \tilde{\delta}_{12} \tag{15}$$

Clearly $\tilde{\delta}_{12}$ is a significant fraction of the initial decoherent total radiant flux as shown in Figure 4(e). We know that

$$\int_{-s}^{s} [2\cos\alpha \, J_1(\beta)/\beta]^2 \, dx = \int_{-s}^{s} 2[J_1(\beta)/\beta]^2 \, dx , \tag{16}$$

and using Eq.s (5-16), the relationship between the coherent and decoherent states, can be expressed as

$$\Phi_{12} = \Phi'_{12} = \tilde{\Phi}'_{12} + \tilde{\delta}_{12} . \tag{17}$$

Eq. (17) simply restates the fact that for the *coherent* state, the presence of the wire makes no significant difference in the total radiant flux entering the lens ($\Phi_{12} = \Phi'_{12}$), and that it is the *same* as in the case when there is no wire present. This leads to the conclusion that the total radiant flux of the pinhole images 1′ and 2′ are not affected by the presence of the wire, if the light is in a *coherent* state at $\sigma_1$. In contrast, the same cannot be said about the decoherent state, since in this case the presence of the wire leads to a loss of $\tilde{\delta}_1 = \tilde{\delta}_2 = \tilde{\delta}_{12}/2$ in the total radiant flux of each image.

### 3.4 Impossibility of interaction/attenuation-free diffraction by an opaque obstacle according to QM

In the discussion of diffraction, textbooks often fail to mention that the initial wavefunction is *always* attenuated after interaction with the opaque obstacle which produces the diffraction pattern in the transmitted wave function perhaps because the relative intensities within a distribution is of interest and thus normalization is justified. An optically opaque obstacle is an impenetrable barrier which has a cross section $e \gg \lambda$. The interaction of a wave function with such an obstacle is a completely *local* process governed by Schrödinger equation, for which a *non-zero* amplitude must be present at the surface of the obstacle. Figures 5(a-c) depict the quantum mechanical simulation of a Gaussian wave packet directly hitting an obstacle (here $e = 30\lambda$) and consequently being partly reflected backwards, and partly diffracted in the direction of initial motion. In our simulation, the obstacle satisfies the Dirichlet boundary condition and is assumed to be a perfect mirror, reflecting the incident wave function without any damping.[26] It is clear that the transmitted part of the wave function is greatly attenuated and contains the telltale diffraction "lobes", enclosed within the dashed ellipse in Figure 5(c).

In contrast, Figures 5(d-f) show the same initial wave packet nearly missing the obstacle. In this scenario, the wave function interacts with only the lower surface of the obstacle, and therefore the reflected and diffracted portions of the wave function are dramatically reduced.

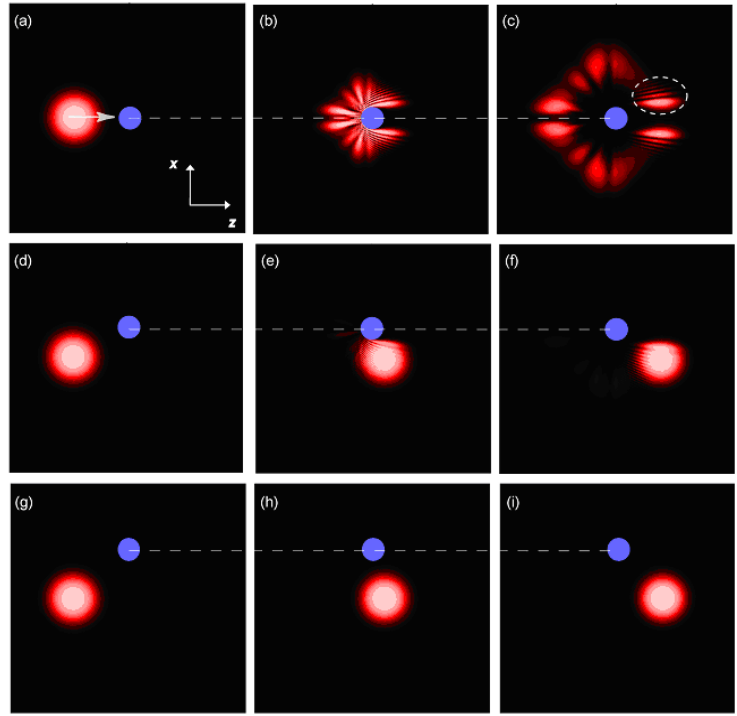

**Figure 5.** Theoretical simulation of the quantum-mechanical effect of an opaque obstacle on the evolution of a Gaussian wave packet for three different positions of the obstacle. (a-c) The wave packet directly hits the obstacle, producing significant attenuation and diffraction in the transmitted light. (d-f) The wave packet interacts with only the lower surface of the obstacle. (g-i) The wave packet nearly misses the obstacle.

Finally, Figures 5(g-h) depict the same initial wave packet, this time completely missing the obstacle. It is clear that the wave function continues to move *undisturbed*, and no diffraction takes place. This is essentially a unitary time development during which the norm of the wave function remains unchanged. Therefore, we can make the following statement: ***If a wave function is not attenuated after passing a region within which a fully opaque obstacle is placed, it is not diffracted by the obstacle, and vice-versa: attenuation ⇔ diffraction.***

### 3.5 Formal proof of interference

Now we shall proceed to formally discuss the condition in which the incident wave function has a large enough lateral extent along the *x*-axis to completely cover the obstacle, yet after passing the obstacle, it is not attenuated (see Figure 6.) We show that: ***the lack of attenuation of the transmitted wave function is a necessary and sufficient condition for the existence of destructive interference at the position of the obstacle.***

**Theorem 1.** Suppose an *apodized* wave function $\psi(x,z,t_1)$ localized along the *x*-axis within $-s \leq x \leq s$ (see Appendix A) is immediately incident on an *opaque* obstacle of thickness $e \gg \lambda$ placed at position $x=u$, $-s \leq u \leq s$. Immediately after the obstacle, the transmitted wave function $\psi'(x,z,t_2)$ continues to move along the *z*-axis. The following relation holds:

$$\|\psi\|^2 = \|\psi'\|^2 \neq 0 \Leftrightarrow \delta = \int_{x_1}^{x_2} |\psi|^2 \, dx = 0 \quad , \qquad (18)$$

where $x_1 = (u - e)/2$ and $x_2 = (u + e)/2$.

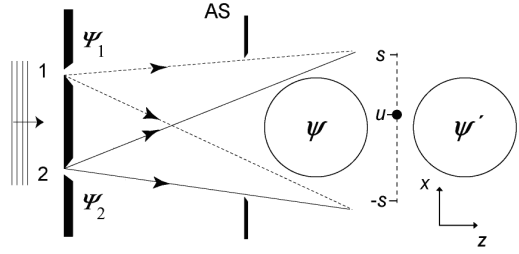

**Figure 6** Apodized wave function $\psi$ moving along the z-axis impinges upon an opaque obstacle placed at $x = u$. The transmitted wave function $\psi'$ would have the same norm as $\psi$, if and only if there is a destructive interference at $x = u$, establishing the presence of an interference pattern.

**Proof.** The interaction of $\psi$ with the obstacle can be written as

$$|\psi\rangle \otimes |\varphi\rangle \xrightarrow{T} |\psi'\rangle , \qquad (19)$$

where $|\varphi\rangle$ represents the obstacle, and $T$ is the unitary time development operator.

We know that $\|\psi\|^2 = \|\psi'\|^2 \neq 0$, therefore

$$\Phi_{12} = \Phi'_{12} > 0 . \qquad (20)$$

But according to Eq. (11) we have $\Phi_{12} = \Phi'_{12} + \delta_{12} > 0$. Therefore, we have

$$\delta = \int_{x_1}^{x_2} |\psi|^2 \, dx = 0 . \qquad (21)$$

**Theorem 2.** For any wavefunction $\psi(x)$, and a given value $x=u$ the following holds:

$$|\psi(u)|^2 = 0 \Leftrightarrow \psi(u) = 0 . \qquad (22)$$

**Proof.** Since $\psi$ is a complex wave function, we have for any given point within the wavefunction a complex vector $\psi(u) = A\, e^{i\theta}$, where $A$ is the modulus of the complex number $\psi(u)$. Since $|\psi(u)|^2 = A^2 = 0$ therefore $A=0$, which necessarily leads to $\psi(u) = 0$. Therefore $|\psi(u)|^2 = 0 \Leftrightarrow \psi(u) = 0$.

**Theorem 3.** For any wave function $\psi(x,y)$, and a given value $x=u$, and $y=v$, the following holds:

$$\|\psi(x,y)\|^2 > 0 \wedge \psi(u,v) = 0 \Rightarrow \psi(u,v) = \psi_1(u,v) + \psi_2(u,v) = 0 . \qquad (23)$$

**Proof.** The wave function has a nonzero norm, and the particular complex vector for a point within the wave function is given as $\psi(u,v) = A\, e^{i\theta} = 0$.

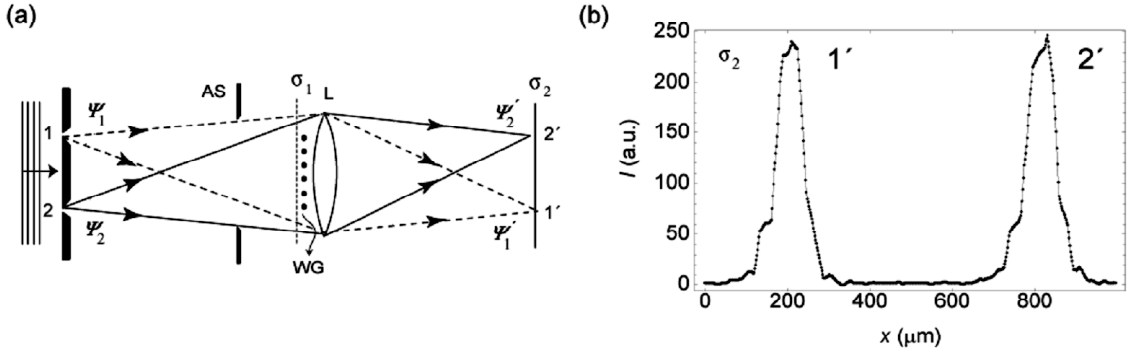

**Figure 7** (a) The configuration testing the effect of the wires in the wire grid (WG). (b) Data representing the images of pinholes 1 and 2. No reduction in the resolution of the images is found at the image plane $\sigma_2$. This implies that no diffraction is produced by the WG and thus WWI is still complete (see text for theoretical justification) so that $K=1$.

$A$ can be written as $A = \sqrt{B^2 + C^2}$, $B = B_1 + B_2 = 0$ and $C = C_1 + C_2 = 0$, $B_n \wedge C_n \neq 0$. We can thus construct at least two complex numbers $\psi_1(u,v) = \sqrt{B_1^2 + C_1^2}\, e^{i\theta} \neq 0$, $\psi_2(u,v) = \sqrt{B_2^2 + C_2^2}\, e^{i\theta} = -\sqrt{B_1^2 + C_1^2}\, e^{i\theta} \neq 0$. It is clear that the sum of these two nonzero complex vectors can be written as $\psi(u,v) = \psi_1(u,v) + \psi_2(u,v) = 0$, which is the superposition of two complex vectors with a phase difference of $\pi$.

## 4. EXPERIMENTAL TEST OF COMPLEMENTARITY

### 4.1 Experimental verification of a nondestructive measurement: methodology

Now that (hopefully) we are theoretically motivated, let us get back to that most important tool of a physicist's trade, the experiment. Figure 7(a) depicts the essential parts of a configuration that can test the validity of PC. In this experiment, we use the absence of photons at the dark fringes (due to total destructive interference), *as opposed to* their arrival at bright fringes (due to total constructive interference), as an equally valid evidence for the coherent wave-like behavior. In order to increase the "shadowing" effect of the wire, we place a series of six equidistant, and parallel thin wires (shown as black dots in the cross-section view of the setup) of thickness $e = 127$ μm ≈ $0.1u$ ≈ $200\lambda$ in front of the lens, at previously measured positions depicted by the asterisks in Figure 1(b), corresponding to the minima of the six most central dark fringes. Each wire is independently placed at the middle of the selected dark fringe with an alignment and positional accuracy of ±1.6 μm. These wires can be considered as a wire grid (WG) with the same periodicity as the IP.

Figure 7(b) shows the irradiance profile of the images at $\sigma_2$, while the WG is present. A comparison with the data in Figure 2(b) immediately demonstrates that the presence of the WG has not affected either the resolution, or the total radiant flux of the images.

The placement of the CCD directly at $\sigma_2$, leads to relatively large errors in the total radiant flux measurement. This is because the diameter of each pinhole image is quite small and few CCD elements receive the incident light, leading to saturation and blooming into the nearby pixels. In order to increase the accuracy, we used the configuration shown in Figures 8(a-c), where mirrors placed at the image plane $\sigma_2$, further separate the incident beams from each pinhole and direct them into different high resolution CCDs 1m away from the image. Naturally, this reflected beam is distributed over a larger number of CCD elements, reducing the local irradiance and thus avoiding the blooming-related errors.

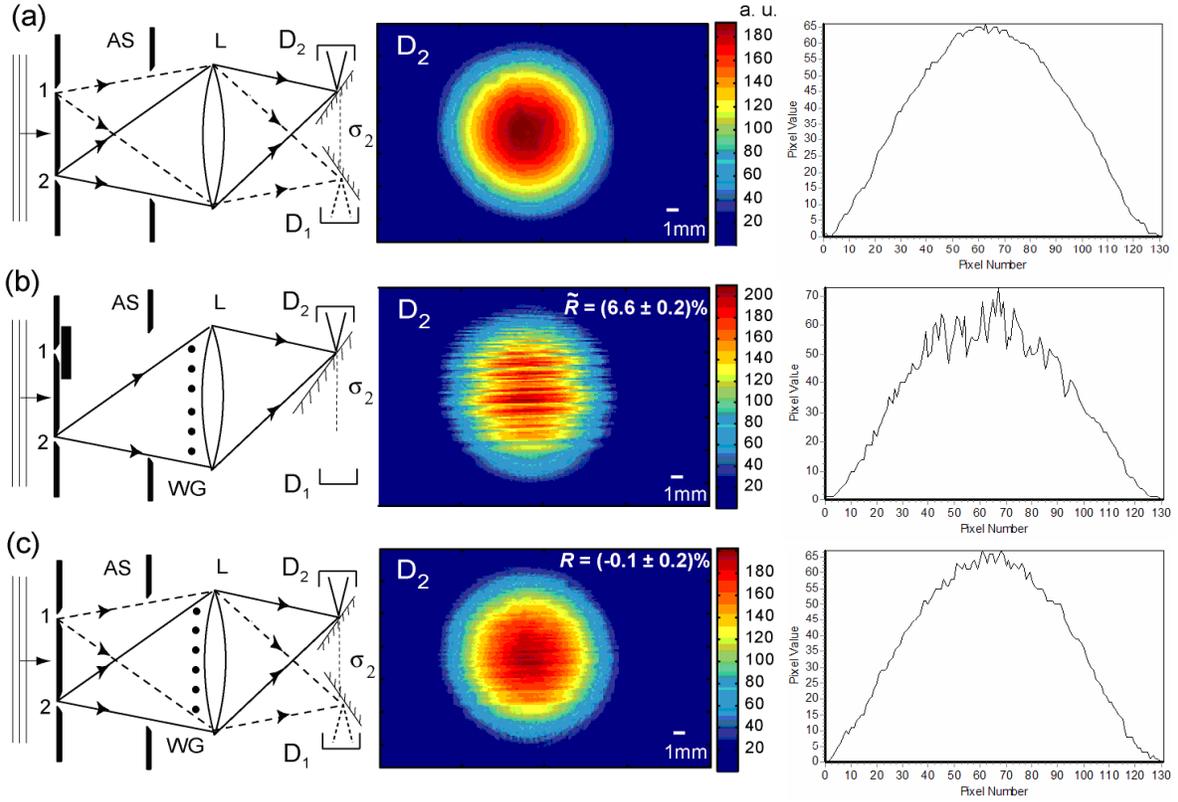

**Figure 8.** Test of Complementarity (a) Control configuration, with both pinholes open and no WG in place. The light from image 2′ is directed to detector $D_2$. (b) Simulation of decoherent state at $\sigma_1$ is achieved by closing pinhole 1, and placing the WG in the path of $\psi_2$. The total radiant flux is reduced by $\tilde{R}$ = (6.6± 0.2)%. Compared to control data the loss of resolution of the image due to diffraction caused by the WG is clear. (c) Both pinholes are open and WG is placed at the dark fringes of the IP. The attenuation of the radiant flux of 2′ is found to be $R$ = (-0.1± 0.2)%, which is negligible. Also the resolution of the image is only slightly reduced compared to control, since no diffraction takes place by WG. Here in violation of PC, *V*=1, and *K*=1, in the same experimental configuration.

Figure 8(a) depicts the control run, where no WG is present and both pinholes are open. The total radiant flux $\Phi_C$ of this run for image 2′ is used to normalize the measurements in the next two experiments. Figure 8(b) shows the configuration and data for the simulation of a *decoherent* distribution of light at $\sigma_1$. One of the pinholes is closed and therefore there would be incident photons on the WG, which attenuates and diffracts the transmitted light gathered by detector $D_2$. Using Eq.s (14) and (15), the normalized reduction in the total radiant flux of image of pinhole 2 for the *decoherent* case is given by

$$\tilde{R} = 100\, \tilde{\delta}_2 / \Phi_C .\qquad(24)$$

The loss of the radiant flux due to the WG in this case is theoretically calculated to be $\tilde{\delta}_2 = \Sigma_6 \tilde{\delta}_2' = \Sigma_6 \tilde{\delta}_{12}/2 = 6.5\%$ of $\Phi_C$. The normalized radiant flux blocked by the wires is found to be $\tilde{R}$ = (6.6± 0.2)% by the analysis of the data, which matches the above theoretical value very well. Also, as expected, it is evident from the density plot of the $D_2$ output that the *resolution* of image 2′ has been significantly *reduced* in comparison to that of the control case.

**4.2 Test of PC**

In similar fashion to Eq. (24), using Eq.s (11) and (12), the normalized reduction in the total radiant flux of image of pinhole 2 for the *coherent* case is given by

$$R_{Coherent} = \delta_2 / \Phi_C. \qquad (25)$$

Figure 8(c) shows the configuration in which both pinholes are open, and the WG is present. The data show that the attenuation of the transmitted light in this case is negligible, $R = (-0.1 \pm 0.2)\%$ indicating that the WG has not absorbed or reflected a measurable amount of light within the margin of error, thus establishing the presence of dark fringes at $\sigma_1$, so that $V=1$. It is also evident that the loss of the resolution of the image compared to the decoherent case is negligible. There is a very good agreement between the theoretical value of $R_{Coherent}=0$ and the observed value $R$. This is compelling evidence for the presence of a perfectly visible IP ($V=1$) just upstream of WG.

## 5. DISCUSSION AND CONCLUSION

Using Eq. (24) and the observed value for $R$, we can define a new parameter:

$$\eta = \frac{\widetilde{R} - R}{\widetilde{R} + R}, \quad 0 \leq \eta \leq 1. \qquad (26)$$

If PC is correct, then in any experiment, we *must* find $\eta = 0$ since the observed value for $R$ must be that of the decoherent case $\widetilde{R}$, due to the fact that we find no reduction in the resolution of the images as shown in Figure 7(b), so that $K=1$. The presence of a perfect IP, would result in a $R=0$, and therefore would lead to an ideal result of $\eta=1$. Bearing in mind the margins of error in our measurements, in this experiment we find that $0.97 \leq \eta \leq 1.1$, again confirming a clear violation of PC. It is expected that this result can be improved upon by reducing the thickness $e$ of the wires in the WG, yet maintaining the condition for opacity ($e \gg \lambda$), and increasing the resolution and sensitivity of the CCDs.

I have endeavoured here to introduce a novel, non-destructive measurement process for the visibility of the IP which can be generalized to any ensemble property, be it spatial, temporal, or otherwise. In the last experiment shown in Figure 8(c), no attenuation of the transmitted light, and no significant reduction in the resolution of the image of pinhole 2 (it could as well have been pinhole 1) is found, although the WG is present in the path of the light. ***It is concluded therefore, that the coherent superposition state at the IP plane $\sigma_1$ persists (V=1) regardless of the fact that the WWI is obtained (K=1) at the image plane $\sigma_2$ in the same experiment.***

One might be tempted to argue that the reliability of the WWI is lost due to the presence of the WG. However, as discussed at length in sections 3.4 and 3.5, since the diffraction by WG could be the only reason for the reduction of $K$, we have established no such diffraction takes place, since no attenuation in the transmitted light is observed. *This simply means there was no light incident on the wires in the WG to diffract.* Therefore, since no diffraction takes place, no reduction in $K$ is possible. Thus it is established that in the *same* experiment, sharp complementary wave and particle behaviors can coexist so that $V^2 + K^2 \approx 2 > 1$, violating Eq. (1) and the PC.

It is worth mentioning that since the so-called "delayed-choice" class of experiments[23] rely primarily on the validity PC, the results of this experiment demonstrate that there is really no "choice" to be made, as the coherent superposition state remains intact although WWI is obtained. Since the arguments presented in this paper are valid for all quantum particles, it is plausible that equivalent experiments could be performed involving electrons or neutrons with identical results to this experiment.

# COROLLARY

Since the initial results of the experiment were made available publicly in March of 2004,[27] numerous critiques of the interpretation of the experiment were offered by the physics community. It would be impossible to discuss all those criticisms due to the page limitation of this publication, however, I would like to suggest two new experiments which may a go a long way in answering most of the critics.

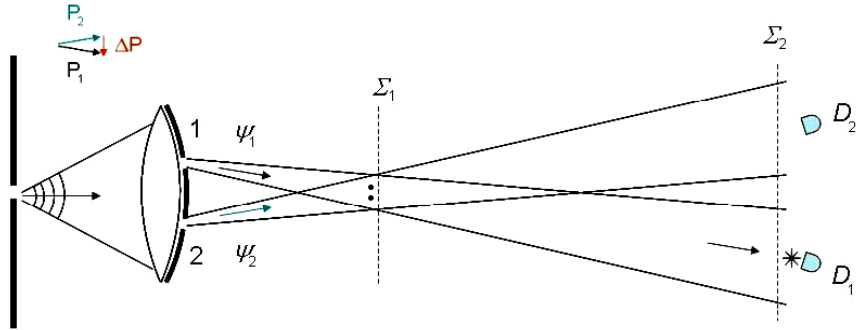

**Figure 9.** The configuration for the first suggested experiment.

The first suggested experiment is a modified version of Wheeler's original delayed-choice experiment, in which two mutually coherent beams simply cross each other. Figure 9 depicts two beams crossing each other at plane $\Sigma_1$ and unitarily evolving unto well-separated beams further downstream at $\Sigma_1$. It is clear that at $\Sigma_1$ the beams will interfere and by the passive placement of the wires at the minima we can gain information about the visibility of the interference there. A single-photon detector, say $D_1$ registers a photon in $\Sigma_2$. Since the linear momentum of the photon is conserved, we cannot accept the proposition that this photon could have originated in pinhole 2, due to the fact that it must have changed its direction of motion at some point. We know that the wires cannot exchange momentum with the photon since they do not intercept it, and thus complete WWI is obtained, thus violating PC again.

The second experiment is based on the assumption that PC is indeed violated. The take-home message of such a violation is that the so-called collapse of the wavefunction does not take place. If so, the question is whether "empty waves" could help produce interference at the last beam splitter in a Mach-Zehnder type interferometer. This experiment is a modified version of the empty wave experiment of Mandel *et al.*[28] conducted in 1991 to investigate whether empty waves can induce coherence. The pump laser is incident on a beam splitter and equally irradiates two identical down-conversion crystals NL1, and NL2. The idler beam from NL1 is aligned such that its optical path overlaps with the idler beam from NL2. The signal beams from both crystals are brought together before detector $D_s$ and a first order IP with visibility of about 33% is obtained. Now, I modify their experiment in two critical ways: (1) allow all of $i_1$ to enter NL2 to ensure maximum induced coherence. (2) place two identical 50-50 beam splitters $BS_1$ and $BS_2$ just before the final beam splitter. Step (2) gives us the opportunity to investigate the effects of the wavefunction collapse by observing say the upper beam before (A), at (B), and after (C) detection of a photon at $D_s$. This means we can now compare the resulting first order spatial IP at $D_s$ with and without the beam splitters and with and without the collapse of the wavefunction for $s_1$. If we observe no reduction in the visibility of the IP (given we allow the same number of photons to accumulate), then we can at least claim that the empty waves are capable of guiding a real photon to allow it to participate in an IP.

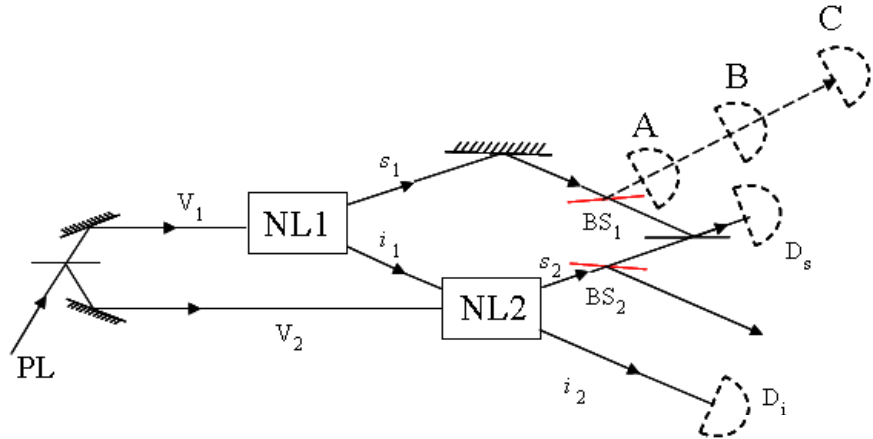

**Figure 10.** The configuration for the second suggested experiment.

Should this second experiment prove positive, the next step would be to isolate the empty waves and observe their dynamical properties by perhaps accumulating large numbers of such waves within a carefully controlled optical cavity and looking for any changes in its temperature. Figure 10 depicts a possible setup. The isolation is achieved by opening a delayed Optical Gate (OG)--eg. a Pockels Cell, only after detector D has detected the single photon emerging from the beam splitter. From the point of view of quantum mechanics, upon such detection, the wavefuntion should collapse, and the other channel must be considered as completely empty. If we observe any physical properties for this beam, we will have discovered a new form of electromagnetic field and would have to revise all our theories of radiation and detection.

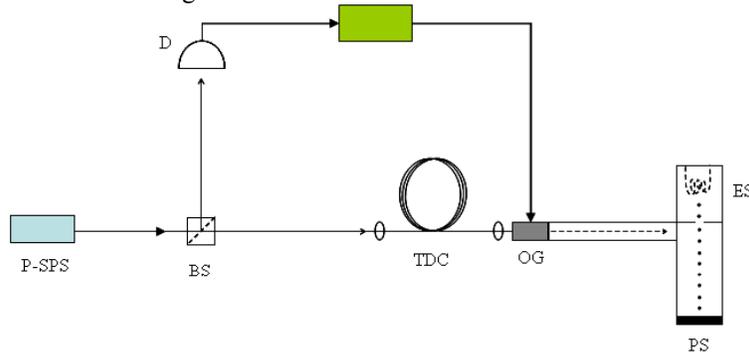

**Figure 11.** The configuration for the third suggested experiment.

## ACKNOWLEDGEMENTS
I would like to thank G. B. Davis, D. W. Glazer, J. Grantham and other financial supporters of this research, to all of whom I am deeply grateful. I also would like to thank Christopher Stubbs and Harvard University Department of Physics for their hospitality and the opportunity to duplicate and verify the experiment, which was initially conducted at IRIMS, as well as Eduardo Flores and Ernst Knoesel for many delightful conversations and Rowan University Department of Physics and Astronomy for their ready assistance and support.

# APPENDIX A

The total probability of finding a photon with wave function $\Psi(x,y,z,t)$ somewhere in space is given by

$$\|\Psi(x,y,z,t)\|^2 = \int_{-\infty}^{\infty}\int_{-\infty}^{\infty}\int_{-\infty}^{\infty} |\Psi(x,y,z,t)|^2 \, dx \, dy \, dz \quad. \tag{A.1}$$

In this Letter, we use the *one-dimensional* notation $\Psi(x)$ for simplicity of argument without any loss of generality and use the equivalence of the classical notion of irradiance and quantum mechanical probability distribution such that we have

$$\Phi = \|\Psi(x)\|^2 = \int_{-\infty}^{\infty} |\Psi(x)|^2 \, dx = \int_{-\infty}^{\infty} I(x) \, dx \quad, \tag{A.2}$$

where $\Phi$ is the total radiant flux, and $I(x)$ is the classical irradiance at position $x$. Due to the practical impossibility of scanning the entire space, we employ apodization in our experiment for the wave functions $\Psi_1$ and $\Psi_2$ so that only the maximal Airy disks are allowed to go through the aperture stop AS and the resulting apodized wave functions $\psi_1$ and $\psi_2$ emerge. These wave functions are bounded within $-s \leq x \leq s$, where $s = 3.833 \, l \, \lambda/b$, $l$ is the distance of plane $\sigma_1$ from the dual pinhole, and $b$ is the diameter of each pinhole [25]. Therefore, we have

$$\psi_i(x) = \begin{cases} 0 & \text{for } x > s \\ \Psi_i(x) & \text{for } -s \leq x \leq s \\ 0 & \text{for } x < -s \end{cases} \tag{A.3}$$

where $i=1,2$.

Bearing in mind that both $\Psi$ and the irradiance $I$ are functions of $x$, for apodized wave functions, the total radiant flux in (A.2) is reduced to

$$\Phi_i = \|\psi_i\|^2 = \int_{-s}^{s} |\psi_i|^2 \, dx = \int_{-s}^{s} I_i \, dx \quad. \tag{A.4}$$

*afshar@rowan.edu, phone: (856) 256-4859, fax: (856) 256-4478; http://users.rowan.edu/~afshar/